To appear in:

# Theory in the Pacific, the Pacific in Theory: Archaeological Perspectives

Editor: Timothy Thomas

Routledge Press

Chapter Title:

Evolutionary Theory in Pacific Island Archaeology

Ethan E. Cochrane, The University of Auckland

## Introduction

The Pacific Islands have been a crucible for the development of evolutionary concepts in archaeology and anthropology that are now used world-wide. Pacific researchers' experimentation with evolution (e.g., Sahlins, 1960) was originally based upon the cultural evolution of nineteenth century scholars, and their ladder of evolutionary progress. Since the 1970s, however, and under some influence from Americanist New Archaeology (O'brien, Lyman and Schiffer, 2005), evolutionary research in the region dismissed the more strident, progressivist Victorian ideas and focused increasingly on adaptive processes that might explain the functioning of past societies (e.g., Kirch, 1980). Later still, another evolutionary framework in the Pacific developed from a mixture of mathematical models of cultural trait transmission (e.g., Cavalli-Sforza and Feldman, 1981; Boyd and Richerson, 1985) and a generalisation of evolutionary concepts beyond biology (e.g., Dunnell, 1978).

This chapter outlines the use of evolutionary theory in Pacific Island archaeology, noting the points of commonality and divergences of thought in the most prominent evolutionary frameworks. The next section reviews these frameworks in the Pacific and focuses on their most important theoretical contributions. The earliest framework, the Ladder of Cultural Evolution, while tinged with dubious notions of progress, emphasised the importance of phylogenetic history or cultural relatedness, and encouraged theorizing on processes of societal change. A framework focused on phylogenetic history and adaptation emerged from ladder-like approaches, but developed and expanded both a range of adaptive processes to account for the similarities and differences amongst populations and sophisticated techniques to generate hypotheses of cultural or phylogenetic relatedness. A third framework, focused on selection, drift and related artefact classifications privileges explanations that invoke evolutionary processes, complementary to proximate processes. Proximate processes refer to the socio-natural dynamics that generate the costs and benefits associated with variants of a behavioural type, while evolutionary processes refer to the sorting mechanisms, such as selection, that act on those costs and benefits to generate behavioural distributions through time and space (Winterhalder and Smith, 2000, pp. 52-54).

The subsequent section compares two frameworks' explanation of a pivotal event, the Lapita migration from Near to Remote Oceania. This comparison demonstrates the application of proximate and evolutionary mechanisms and the recognition of phylogeny, but also proposes that robust explanations should employ descriptions of the archaeological record using units, artefact types or other classes, that are explicitly linked to evolutionary theory.

4**Evolutionary Theories in the Pacific**

Like many attempts at intellectual categorisation, the following division of evolutionary frameworks in Pacific Island archaeology is meant to highlight distinguishing features, but of course these frameworks are sometimes combined in single analyses or bodies of work.

*The Ladder of Cultural Evolution: Stages and Adaptive Advances*

The evolutionary concepts first used to explain variation in Pacific Island material culture and practices derive from a ladder-like view of cultural evolution. In this view populations occupy different rungs on a ladder of increasing complexity or progress. For example, d'Urville's (1832) widely-used division of the Pacific into Melanesia, Micronesia, and Polynesia included an assumption that the people in these geographic regions were of differing social and political complexities, with Polynesians the most advanced and Melanesians the least (Clark, 2003, pp. 157-158). However, the idea that populations of different complexity were evidence of processes of evolutionary development was not elaborated for several more decades until Spencer (1855, 1857, 1873) prominently argued that human social development, from lower forms to higher forms, was caused by the cultural inheritance of human inventions and a force leading to perfection or progress (Freeman, 1974). Rungs on the ladder of progress were also given labels including Morgan's (1877) savagery, barbarism, and civilization that fit with the age-system of European archaeology popular at the time (Dunnell, 1988).

The ladder-like view of cultural evolution is distinct from Darwinian evolution (Figure 1) and is not built upon the latter's theoretical assumptions of variation, transmission and selection (Blute, 1979; Dunnell, 1980; Mesoudi, 2016). Instead, and despite the use of phrases such as 'survival of the fittest' by 19th century cultural evolutionists, the ladder-like view of cultural evolution is transformational, and not based on a Darwinian selection process (contra Carneiro, 2003, pp. 68-73). In the transformational view explanations state the processes by which one societal type transforms to another. Tylor (1870), as an example of the time, suggested technology transfer might explain the transformation of a tribe, a low rung on the ladder, to a higher civilization. Even with more rungs or finer-grained types—incipient and simple chiefdoms (Steponaitis, 1978; Earle, 1987), archaic and modern states (Marcus and Feinman, 1998)—the explanatory focus remains the processes whereby one finer-grained type transforms to another. In the transformational view, societal types exist in the sense that explanations account for how types originate.

<Figure 1 about here>

This contrasts with some explanation in Darwinian evolution where types do not exist independent of the analyst, but are conceptualized and constructed as measurement units (Mayr, 1959), the artificiality of which is obvious if you imagine a type in the human lineage, an *H. erectus*, birthing an *H. sapiens*. Whit a similar approach in archaeology, types are constructed and used to characterise variant distributions in the archaeological record (Cochrane, 2001) and it is these variant distributions that are the focus of explanation (Figure 1). As types are a product of archaeological classification procedures, their origin is a non-question.

The influence of 19[th] century cultural evolutionism on Americanist archaeology and ethnology was dimmed by Boas's focus (e.g., 1911) on the particularistic aspects of cultures[1].

---

[1] Late 19[th] and early 20[th] century British anthropologists in the Pacific were influenced by evolutionary thought (see e.g., Kuklick, 1996) but their influence on modern archaeology (post-WWII) in the region has been more limited.

By the mid-20th century, however, ladder-like cultural evolution was re-kindled by Leslie White who linked Spencer's nebulously defined force of progress to specific empirical variation across populations. White (1959, pp. 144-145) clearly associated cultural evolution with the amount of energy extracted from the environment by populations:

> Social systems are but the social form of expression of technological control over the forces of nature. Social evolution is therefore a function of technological development. Social systems evolve as the amount of energy harnessed per capita per year increases, other factors remaining constant. This is to say, they become more differentiated structurally, more specialized functionally, and as a consequence of differentiation and specialization, special mechanisms of integration and regulation are developed (see also White, 1943).

Sahlins described White's proposal as general evolution, the focus of which is "the determination and explanation of the successive transformations of culture through its several stages of overall progress" (1960, p. 29). General evolution, Sahlins also argued, was distinct from specific evolution, which was concerned with adaptation within cultural lineages, for example, how the diversification of production systems in related Polynesian chiefdoms show adaptive advances (Sahlins, 1958). Specific evolution was also related to Steward's (1955) multilinear cultural evolution that investigated similarities between cultural traditions to identify adaptive regularities (e.g., patrilineal bands), particularly those associated with certain environments. In contrast to Steward, however, Sahlins considered a complete explanation of culture to include both specific and general evolution as "the former is a connected, historic sequence of forms, the latter a sequence of stages exemplified by forms of a given order of development" (Sahlins, 1960, p. 33).

Nevertheless, Sahlins (1958) focused on specific evolution in his study of social stratification in Polynesia where he identified adaptive regularities, in large part, by the number of status levels present, and grouped them into three types (from greater to fewer status levels): ramage systems, descent-line systems, and atoll systems. Since it was not possible to generate data on productivity or per capita energy extraction in these societies (as White's proposal required), Sahlins argued that variation in surplus food production, measured by the spatial scale and frequency of its distribution, is a valid proxy for per capita energy extraction (Sahlins, 1958, pp. 107-110, Table 1). He concluded that societies with greater social stratification, with ramage systems for example, had greater surplus, and therefore greater per capita energy extraction, confirming White's general evolutionary thesis.

White's notion that more evolved societies should also exhibit relatively greater structural differentiation and functional specialisation was also confirmed by Sahlins and others (e.g., Service, 1962) working in the Pacific. For example, Sahlins described how more highly developed chiefdoms should have a greater number of cultural sub-systems and more specialised sub-systems than less developed tribes (Sahlins, 1960, pp. 21-22 and 35-37; 1972).

While mid-century cultural evolutionists correlated societal stages with productivity measured in various ways, the processes that cause societies to become more productive and transform from one rung on the ladder to the next have been continuously, if somewhat infrequently, debated (Carneiro, 2003, pp. 155-156). These processes, such as Goldman's (1955) inherent status rivalry, Friedman's (1981) structuralist transformation, Kirch's (1984) competition in limited environments, Earle's (1997) political economy, and the culturally specific rules or social logic of Flannery and Marcus (2012), all include individuals, groups or societies adapting to new socio-natural environments. Current archaeological research that privileges these processes relies, at least in part, upon emic perspectives or human motivations to explain cultural evolution and changes in cultural complexity (e.g., Kirch, 2010; Lepofsky and Kahn, 2011; Clark, 2017; Quintus and Cochrane, 2018).

Ladder-like cultural evolution has influenced archaeological thinking on the processes that may explain changes in cultural complexity, but Ladder-like cultural evolution is also widely employed in Pacific archaeology in another way. The general characteristics of a rung on the ladder, a societal type, are used to culturally reconstruct a time-space portion of the archaeological record as representing, for example, a tribe or chiefdom. The association of some characteristic of the archaeological record with a particular type – monumental architecture with chiefdoms, for example – is then used to aid in the reconstruction of other aspects of ancient society (e.g., Stevenson, 2002; Wallin and Solsvik, 2010). Clark et al. (2008) for example, suggest that, the cultural chronology of monumental tomb constructions and habour modifications at Lapaha, Tonga, supports a pre-AD 1450 date for marine transport associated with *'inasi*, the first-fruits ceremony found in many Pacific chiefdoms. Tongan chiefs "may have been using the *'inasi* and other ceremonies to promote social cohesion among local and non-local groups in the newly constituted and geographically dispersed chiefdom" (Clark, Burley and Murray, 2008, p. 1006). Other aspects of chiefdom society have been reconstructed at various times and places from the Pacific island archaeological record including alliances between New Caledonia chiefdoms (Sand, Bolé and Ouetcho, 2003), the possible dual sacred and secular leadership roles in ancient Samoa (Wallin and Martinsson-Wallin, 2007), and attached specialist carvers in Rapa Nui (Simpson, Van Tilburg and Dussubieux, 2018)

Leonard and Jones (1987) pointed out a potential problem with this cultural reconstruction approach over 30 years ago after analysing the data within Murdock's (1967) *Ethnographic Atlas*, and focusing on three variables: community organisation, settlement pattern, and class stratification. They demonstrated that the values within these variables that typically define cultural evolutionary stages – for example the segmented communities, complex settlements, and dual stratification of chiefdoms – do not often appear as a package within particular ethnographically-described societies (Leonard and Jones, 1987, Table 1). In other words, using an archaeologically visible trait, such as monumental architecture, to infer additional, less easily observed traits of particular societal types, can lead to unrecognised errors, a problem occasionally recognized in the Pacific (Sand, 2002) and elsewhere (Wengrow and Graeber, 2015; Hildebrand et al., 2018).

*Phylogenetic History and Adaptation: the Legacy of Islands as Laboratories*
In his discussion of specific and general evolution, Sahlins described the study of specific evolution of cultures as the examination of cultural traits "arising through adaptation" (Sahlins, 1960, p. 25) within phylogenetically related societies. According to Sahlins when a group enters a new environment, traits arising through adaptation are synthesized from the group's previous traits, or are diffused from other groups, or both (Sahlins, 1960, p. 24). These ideas were developed in the Pacific most prominently by Kirch (1984) in his seminal book *The Evolution of the Polynesian Chiefdoms* (see also Kirch and Green, 1987; Kirch and Green, 2001). In particular Kirch argued that the culturally-related societies of Polynesia provide an opportunity to investigate how variation in environments, interaction, and demography have influenced traits arising through adaptation in separate, but related societies.

The focus on adaptation is a defining feature of this approach. Adaptation is typically conceptualised as a proximate, ecological process whereby entities, such as human groups, continuously achieve a better fit with their changing environment (cf., Steward, 1955; Kirch, 1980). This privileges systems-like explanations, which are a component of evolutionary explanations in general, and identify how the values within a set of interrelated variables change (e.g., Kirch et al., 2012). These explanations are compatible with selection processes as they focus on the adaptive function of an evolved system, but they do not demonstrate why one system evolved relative to another (Dunnell, 1980). To illustrate the latter question, consider several behaviours explained as adaptive systems that minimize lethal competition between groups over limited resources: Rapa Nui moai construction and costly signalling (Dinapoli et al., 2017); settlements locations in Fiji and economic defendability (Field, 2004); subsistence and flexible territories in the Channel Islands (Kennett and Clifford, 2004). To determine the evolutionary processes that resulted in a particular system – costly signalling and not economic defendability, for example – requires identifying the chronological and spatial history of behavioural alternatives and their relative fitnesses. Although such research has rarely occurred, Allen's (2011) study of different political systems provides a structure for this kind of analysis.

Various conceptual frameworks are used to generate explanations of societies as adaptive systems with Political Economy and Human Behavioural Ecology frameworks particularly prevalent in Pacific research. Political Economy frameworks focus on types of individuals, such as elites, and their variable ability to draw on different sources of power "to strategically direct (and resist) the actions of social groups" (Earle, 1997, p. 208). Earle argues that three sources of power – economic, military, and ideological – are most important for explaining changes in cultural complexity and control in societies. Natural environments and historical circumstances will also shape the use of particular power sources (Earle, 1997). Reviewing the history of Political Economy frameworks, Earle (1997, pp. 68-70) distinguishes those that are "voluntarist, adaptationist theories" from "coercive, political" ones. The former describe societies where elites and non-elites perform different, integrated roles that create an adaptive system, for example chiefly management of collective water-rights for commoner farmers. The latter describe societies where different types of individuals variously compel others, or resist, through use of power. In fact, both kinds of Political Economy frameworks are adaptationist in the sense that the explanations generated consider the behaviour of types of individuals as interrelated with the variables of changing natural and social environment. In Hawaii for example, the political economy of 600 or more years ago was based on control of staple finance (crops) in an area where agricultural productivity was high, and could be increased through means of landesque capital. Contrastingly, in the Thy region of Denmark during the Neolithic-Bronze Age transition, the political economy was based on control of wealth finance (cattle) that could be easily moved and exchanged in a natural environment with little possibility for control of crop-based staple finance. Political Economy frameworks are also adaptationist in that they typical produce system-like explanations and conceptualize culture as comprised of sets of interrelated variables, such that change in one variable may affect others (Figure 2).

<Figure s about here>

Human Behavioural Ecology (HBE) frameworks may also form the foundation for adaptive explanations in the Pacific. In archaeology, HBE frameworks have long-assumed that relevant human variation is patterned by natural selection acting upon the distribution of fitness-related behaviours in particular social-ecological contexts (Irons, 1979; Smith and Winterhalder, 1992; Bird and O'connell, 2006). Two additional assumptions are important: first, behaviours in a population are assumed to be a product of inheritance, but the type of inheritance, genetic or cultural, is not relevant (the phenotypic gambit; Smith and Winterhalder, 1992; Codding and Bird, 2015); second, HBE explanations are typically based on either optimisation (Macarthur and Pianka, 1966) or game-theory methods (Lewontin, 1961). When using optimisation methods, the distribution of individuals' behaviours is predicted to optimize some currency, such as net caloric return of hunted animals. The Prey Choice Model, for example, can be used to explain variable frequencies of hunted animals in a diachronic deposit by calculating the weights of different taxa (e.g., Morrison and Cochrane, 2008). Game-theory methods are used to predict the distribution of behaviours in a population when the fitness of a behaviour is influenced by the probabilities of interactions between different behaviour-types in a population. The Hawk-Dove model, for example, can be used to explain the distribution of aggressive and acquiescent behaviours in a population when the chances of aggressive-aggressive, aggressive-acquiescent, and acquiescent-acquiescent interactions vary. Hayman (2019) has used this model to investigate the changing numbers of competitors for high-status titles in ancient Tonga relative to those who acquiesce at less potential cost, but receive lower ranks (cf. Aswani and Graves, 1998)

HBE, Political Economy and other adaptationist evolutionary research in Pacific archaeology often examines different systems that share phylogenetic history or cultural relatedness (e.g., Kirch, 1990; Allen, 2011; Dinapoli et al., 2017). Phylogenetic history may distinguish analogous or convergent similarity from homologous similarity or similarity that is a product of relatedness (Binford, 1968; Kirch and Green, 1987). Phylogenetic history is typically reconstructed either quantitatively using linguistic (e.g., Gray, Drummond and Greenhill, 2009) or, less commonly, artefact data (e.g., Cochrane, 2015), or comparatively using historical linguistic methods (e.g., Blust, 1995) or artefact types (e.g., Emory, 1946).

For archaeologists, phylogenetic relationships, irrespective of how they are generated, are typically a valuable aid in archaeological interpretations. Gray and Jordan (2000), for example, constructed quantitative linguistic phylogenies to evaluate competing hypotheses concerning Lapita migration while Pawley (2018) used comparative linguistic methods to propose a particular settlement history for Fiji-West Polynesia. While archaeological interpretation of various phylogenies has proven popular (e.g., Green, 1966; Kirch and Green, 1987; Pawley, 1996; Sheppard and Walter, 2006; Burley, 2013) there are two frequently voiced concerns (Dewar, 1995; Terrell, Hunt and Gosden, 1997; Moore, 2001). One is the degree of isomorphism between language phylogenies and the phylogenetic relationships defined by human biology or artefact traditions. This amounts to the significant argument that language, culture, and biology might not travel together down the same evolutionary pathways (see comments in Bedford et al., 2018). The other concern is that the branching relationships required by many phylogenetic methods might be inappropriate models to investigate cultural change (Collard, Shennan and Tehrani, 2006). One remedy for the first concern is to construct phylogenies based on different realms of human variation, languages and artefacts for example, and identify mismatches that require explanation (Tehrani, Collard and Shennan, 2010). The concern that branching models of relatedness are inappropriate can be approached as an empirical problem with branching and reticulate models applied to the same data (e.g., Cochrane and Lipo, 2010; Shennan, Crema and Kerig, 2015).

Phylogenetic relationships are also investigated to determine if similarities in different populations are homologous, such as shared ancestral traits, or if similarities are analogous, arising due to independent invention, among other possible processes. Distinguishing between the two is known as Galton's problem (Mace and Pagel, 1994) and has been investigated in the Pacific with reference to agriculture techniques, as well as fishhooks (e.g.,

Allen, 1992; Pfeffer, 2001), and ceramics (e.g., Winter et al., 2012). Considering agricultural techniques, the processes that explain similarities in the origins and distribution of raised beds (Kirch and Lepofsky, 1993), simple flooding (Kuhlken, 1994, p. 364; Mccoy and Graves, 2010:94), windbreak farming (Mccoy and Graves, 2010:94), and lithic mulch cultivation (Barber, 2010, p. 76) are all debated with specific proposals based variably on reconstructed proto-lexemes, archaeological chronologies and distances between island groups.

Reconstructing the cultural characteristics of ancestral societies at the branch points or nodes of cultural phylogenies is another focus of phylogenetic history and adaptation research. This has been pursued by Kirch and Green (1987, 2001) using a triangulation approach whereby controlled comparison across different domains of human variation – ethnography, linguistics, material culture, and biology – is used to develop cultural reconstructions. Beginning with a comparative linguistic phylogeny, they propose that Ancestral Polynesian Society, the population from which East Polynesian groups radiated, was likely situated in Tonga, Samoa, and some of their nearby islands. Using both lexical reconstruction of proto-Polynesian language, and comparative ethnography, Kirch and Green reconstruct, for example, the spatial segregation of domestic activities and argue that separate cooking and dwelling structures existed in the region and time period (i.e., from about 2300 – 1000 BP), although there is currently no archaeological evidence of this (Kirch and Green, 2001, p. 196). A related research program using quantitative approaches to phylogenetic reconstruction has recently emerged. This research uses linguistic trees produced within a Bayesian framework to estimate the likelihood of particular proto-language stages, and concomitantly, the likelihood that certain concepts or social forms represented by reconstructed lexemes were present (e.g., Currie et al., 2010; Kushnick, Gray and Jordan, 2014).

*Selection, Drift and Other Evolutionary Mechanisms*

The most recent evolutionary approach to be applied in the Pacific focuses on mechanisms such as selection, drift, and other cultural transmission processes to explain the distribution of archaeological (e.g., Graves and Cachola-Abad, 1996; O'connor, White and Hunt, 2017) and ethnographic traits (e.g., Rogers and Ehrlich, 2008). This framework, typically called evolutionary archaeology (Shennan, 2008), and sometimes cultural evolution, a label confusingly similar to mid-20$^{th}$ century work, derives from two intellectual lineages. The first is quantitative cultural transmission research beginning about the 1970s (e.g., Cavalli-Sforza and Feldman, 1981; Lumsden and Wilson, 1981; Boyd and Richerson, 1985), a programme later termed dual-inheritance theory or gene-culture co-evolution. The second lineage includes research focused on drift and selection applied to explicitly constructed artefact classes (e.g., Dunnell, 1978, 1992). Both lineages are also inter-woven with human behavioural ecology (Broughton and O'connell, 1999; O'brien and Lyman, 2002).

Evolutionary archaeology research in the Pacific (and elsewhere) regularly uses artefact classes specifically created to investigate mechanisms defined in evolution theory (Graves and Ladefoged, 1995; Cochrane, 2002; Morrison, 2012). These artefact classes are the types by which variation in the archaeological record is quantified to produce variant distributions. Archaeological variant distributions are then compared with the distributional expectations of evolutionary mechanisms to determine if a specific mechanism, such as drift, is an adequate explanation. The comparison of empirical distributions and theoretical expectations facilitates evaluation of artefact classification as the arbitrary decisions in classification must be justified relative to the proposed evolutionary mechanisms (Lipo, 2001; Glatz, Kandler and Steele, 2011). Cochrane (2009), for example, evaluated variously defined Fijian ceramic classes for their ability to track drift processes by examining class criteria against theoretical expectations of continuous distributions and others.

Allen's (1996) work on ancient fishhooks is an excellent example of creating specific artefact classes to investigate evolutionary mechanisms. By using hook types whose definitional criteria are hypothesized to largely measure the results of cultural transmission, Allen was able to show stochastic (i.e., drift) change over time in hook-type abundances from particular islands. After classifying the same hooks with criteria hypothesized to track variation in fishing environments, she demonstrated that the distribution of these hook types in different populations is likely explained by selection and adaptation to different prey types. Applying a similar classification-focused approach to ethnographic variation in canoes, Rogers and Ehrlich (2008) argued that decorative canoe traits such as prow carvings change rapidly, as predicted by drift, and more than functional traits such as outrigger attachment types whose distribution is explained by selection and adaptation.

As artefact classes in evolutionary archaeology are designed to investigate evolutionary mechanisms they will not necessarily match emic or ethnohistoric classes. Allen's (1996) fishhook types, for example, might not be recognised as such by ancient fishers. While these artefact types may or may not have emic reality, their usefulness is evaluated by both justifying the assumptions that underlie their construction, why for example might fishhook head morphology track transmission, and by their ability to produce empirical patterns that can be compared to theoretical expectations. In comparison, other evolutionary frameworks sometimes use observational units proposed to map onto past emically meaningful variation.

For example, Kahn and Kirch's (2011; 2015) Political Economy explanations of variation in Society Islands temple architecture employs culturally relevant types drawn from ethnohistory such as shrines, *marae*, community-level *marae*, particular kinds of priestly dwellings, and structures in which sacred objects were stored. They argue that greater architectural complexity over time and increasing spatial differentiation of these emic architectural types resulted from increasing elite manipulation of ideology and control of staple finance. Cochrane (2015) also examined temple architecture across the Society Islands and Polynesia, but his approach to archaeological classification was based on expectations for the architectural variables whose distribution would be patterned predominantly by transmission and drift. Briefly, these expectations include that frequencies of different values in variables should be independent, they should not be correlated with different environments, and they should vary spatially and temporally. Cochrane's resulting architectural classes do not match emic categories, but the analysis based on them indicates wide-spread transmission of ideas across East Polynesia, a finding supported by other recent evolutionary archaeology and artefact sourcing studies (e.g. Weisler et al., 2016; O'connor, White and Hunt, 2017).

**A Case Study of Two Approaches**

A case study comparing different evolutionary frameworks applied to the Pacific archaeological record will serve, to highlight points of similarity and divergence, and suggest how multiple frameworks might be used to produce better explanations. The archaeological record of initial migration from Near to Remote Oceania has long been a significant research issue in Pacific Island archaeology and two recent evolutionary studies have attempted to explain this population movement: Earle and Spriggs' (2015) Political Economy explanation and Cochrane's (2018) selection-based approach.

Evidence for the initial migration of populations from Near to Remote Oceania (Figure 3) is based on the distribution of Lapita ceramics, an intricately decorated earthenware. These ceramics appear in the Bismarck Archipelago, without local precedent, most likely between 3535-3234 cal BP (95% probability, Rieth and Athens, 2017). After a pause, the ceramics appear with the first human colonists in southwestern Remote Oceania between 3000 cal BP (Sheppard, 2011) for islands in the west like Vanuatu and New Caledonia , and 2750 cal BP for Samoa (Petchey, 2001), the farthest eastern island colonized at the time. These colonists crossed the Near-Remote Oceanic biogeographic boundary that, in part, kept humans and other species confined to the west for approximately 50,000 years (Green, 1991).

<Figure 3 about here>

Before Lapita pottery appears in Near Oceania, populations there had a long history of hunting and gathering, mid-Holocene agriculture and arboriculture, and animal management, as well as tuber and aroid use beginning in the Pleistocene. After the appearance of Lapita ceramics and associated material culture in Near Oceania these subsistence practices continued at new Lapita occupations and at locations with both pre-Lapita and Lapita deposits (Kirch, 1997, pp. 203-205; Lentfer and Torrence, 2007). A suite of domesticated Asian animals also appear in Near Oceania during and after the advent of Lapita pottery.

After voyaging to Remote Oceania, the Lapita ceramic producing populations encountered new plants, animals, and landscapes and the colonists targeted these rich resources (Nagaoka, 1988; Steadman, Plourde and Burley, 2002; Valentin et al., 2010). They also brought crops such as bananas, and taro, that had a multi-millennia history of human use in Near Oceania (Horrocks and Bedford, 2005; Horrocks and Nunn, 2007; Fall, 2010). Some domesticated or commensal Asian animals also accompanied the colonists (Matisoo-Smith, 2007). There is a substantially greater number of Lapita deposits in Remote Oceania compared to Near Oceania (Anderson et al., 2001).

Those groups who left for Remote Oceania initially maintained contact with Near Oceanic populations as evidenced by obsidian transport and the similar, intricate Lapita pottery decorations repeated on locally-made pottery (Dickinson, 2006) in both regions. By approximately 2700 cal BP this movement between populations had largely stopped, both within Remote Oceania and between the regions, as evidenced by the lack of artefact transport, and the replacement of commonly decorated Lapita pottery with different, archipelago-specific styles in Remote Oceania.

*A Political Economy Explanation of Lapita Movement*

Earle and Spriggs (2015) develop a Political Economy explanation of Lapita movement to Remote Oceania focusing on differential access to power within society. They propose that differential access begins with bottlenecks or the "constriction points in commodity chains that offer an aspiring leader the opportunity to limit access, thus creating ownership over resources, technologies, or knowledge" (Earle and Spriggs, 2015, p. 517). Once an aspiring leader has restricted other's access to a resource she or he can more easily extract surplus from the resource to underwrite control of economic, military, or ideological power sources. Earle and Spriggs note that the archaeological application of Political Economy requires the identification of important resources and their constriction points by focusing on the objects which move through ancient economies.

Earle and Spriggs apply this strategy to the Lapita case through an emic approach to unit formation, in particular relying on Pacific ethnography and cross-cultural comparison to determine what archaeological types would correspond to prestige items, Lapita pots, shell ornaments, and bird feathers for example. They argue, however, that the nature of Lapita prestige items, including the widespread Lapita pots, limited their control, with a similar lack of constriction points and opportunities for control of staple finance. In contrast, they point out that knowledge of long-distance voyaging and the ability to colonize new lands, and with new opportunities for trade, was a knowledge-resource that was not available to all and whose distribution and use might be controlled by a minority. If use of this knowledge was a somewhat competitive arena, this, in part, explains Lapita migration "to search out undiscovered islands that could offer easy subsistence, exchangeable products, and a basis for leaders to establish themselves independently" (Earle and Spriggs, 2015, p. 518).

*A Selection-Based Explanation of Lapita Movement*

Cochrane's (2018) explanation of the Lapita migration is different from Earle and Spriggs (2015). Instead of assuming a relevant explanatory process is individuals' use of power to direct and resist groups, Cochrane divides explanations into ultimate-evolutionary and proximate-adaptive types. In the former the relative advantage of cultural traits causes their differential transmission (i.e., replication) over time and across space. That is, trait distributions are explained by selection. The Remote Oceanic Lapita trait or trait-suite includes a focus on hunting and gathering, and long-distance maritime movement, while the Near Oceanic trait-suite include less focus on both hunting and gathering, and long-distance movement. If there is no differential transmission of these trait-suites, then selection is probably not implicated in Lapita movement into Remote Oceania. For proximate-adaptive explanations, adaptation or the changing correlations of traits and environments is a system-level explanation. These correlations can be explained by physiological responses, learned and flexible behaviours or some combination. And such correlations are triggered by other changes in the system, such as environmental change, technological change, or social changes.

To apply this framework, Cochrane develops a method to identify the archaeological signatures of different movement types, such as range-expansion or range shift, with the distribution of these types predicted by both evolutionary and adaptive processes in different environments. Cochrane argues that the Lapita migration is a range-expansion and the greater rate of replication of Lapita trait-suites in Remote Oceania compared to Near Oceania indicates selection. The differential replication of Lapita trait-suites in Near and Remote Oceania can also be explained as a proximate system-like process with changes in both regional weather patterns and canoe technology that changed the costs and benefits of long-distance voyaging. In short, "While climate change and innovation in voyaging technology lowered the cost of movement for Lapita populations, selection processes related to environmental variation and demography explain the relatively greater rate of Lapita deposition in Remote Oceania" (Cochrane, 2018, p. 543). Importantly, there are several empirical observations that would refute or challenge this explanation. If for example, there is not a greater rate of Lapita trait-suite replication in Remote Oceania compared to Near Oceania, then selection is not a valid explanation.

**Conclusion**

The selection-based approach and the Political Economy approach to Lapita migration encapsulate many of the salient characteristics of various evolutionary approaches in the Pacific, highlight the differences, and suggest where they might be combined for mutual benefit. Evolutionary approaches in the Pacific can be broadly divided into adaptationist, system-focused frameworks and selectionist, transmission-focused frameworks. As has been recognized for decades (Mayr, 1961), a complete evolutionary explanation of behaviour, in this case past behaviour fossilised in the archaeological record, should ideally include both kinds of explanations. To exemplify the possibilities, Cochrane's explanation of Lapita migration does not delve into the proximate social triggers that may have kick-started the movement. It is here that a Political Economy and cross-cultural approach might identify more social competitive processes and associated archaeological units by which to investigate the evolution of range-expansion and other movement types. A selection-based approach might identify the evolutionary mechanisms that explain the probable phylogenetic relatedness of populations such as the pre-Lapita ceramic-using groups in Island Southeast Asia, Lapita populations, and the descendant populations of the Pacific whose hierarchical social systems are often the focus of Political Economy research. Such a division of labour mirrors Mesoudi et al.'s (2006) unified approach to the evolutionary study of culture. Here, Political Economy is cast in the role of micro-evolutionary studies with selection-based approaches investigating macro-evolution.

All evolutionary approaches often espouse a concern with scientific explanations, but what this means differs between frameworks. Minimally science uses theory or ideas, to explain the distribution of empirical phenomena, in this case the archaeological record, and potential scientific explanations should be empirically testable. That is, explanations should be constructed in a way that one can envision an empirical observation that would falsify them. All evolutionary approaches should strive for this, lest we construct unverifiable stories.

**Acknowledgements**

This research was partly funded by the Royal Society Te Apārangi Marsden Fund (contract UOA1709). The comments of Thegn Ladefoged, Timothy Thomas, and an anonymous reviewer improved the manuscript.

Figures and captions:

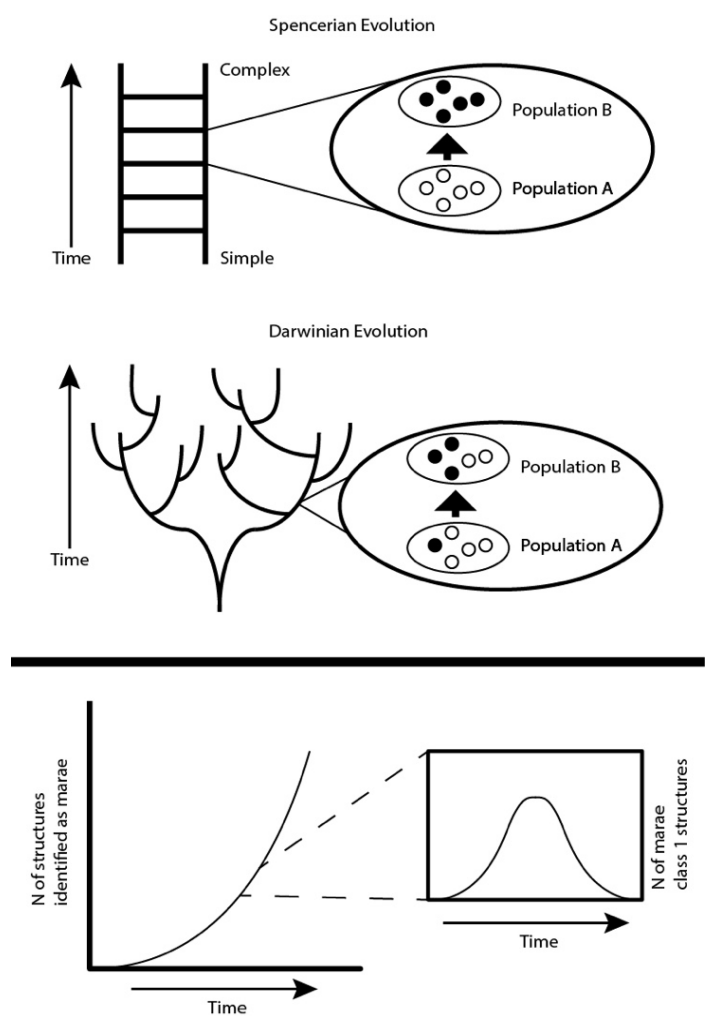

Figure 1. Comparison of Darwinian Evolutionary and Ladder-like Cultural Evolutionary perspectives on change. Top figure, redrafted from Mesoudi (2011, Figure 2.1), demonstrates the transformational approach to variation within types compared to the Darwinian approach to variation within types. Bottom figure redrafted from Cochrane (2001, Figure 10.2) demonstrates how distributions (exponential on left, monotonic on right) produced by archaeologist-constructed cultural trait classes ("marae" and the finer-grained "marae class 1") are the focus of explanation in evolutionary archaeology

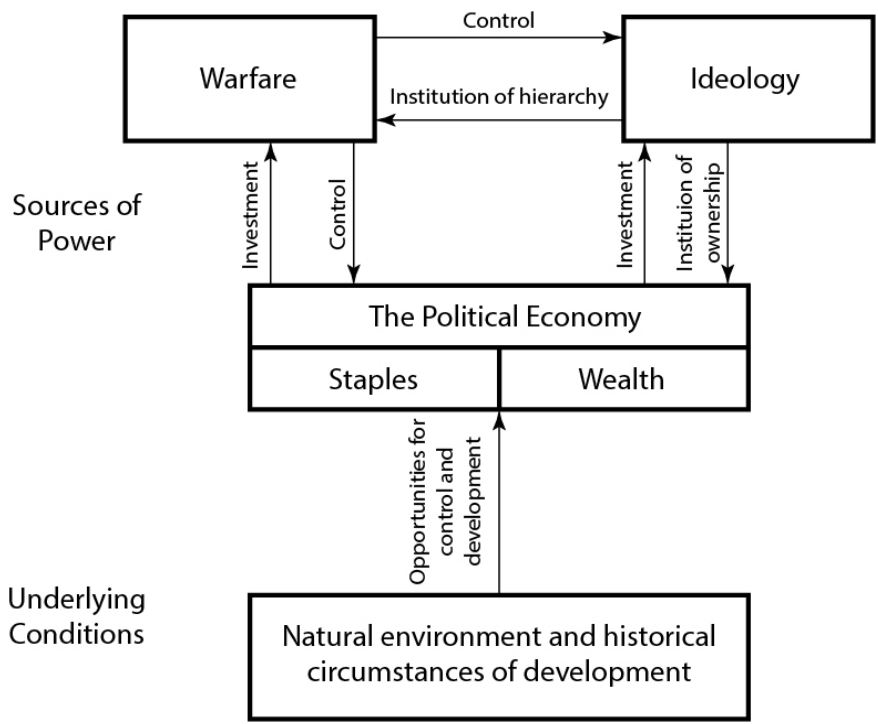

Figure 2. Political Economy explanations as systems of interrelated variables, redrafted from Earle (1997, Figure 6.1).

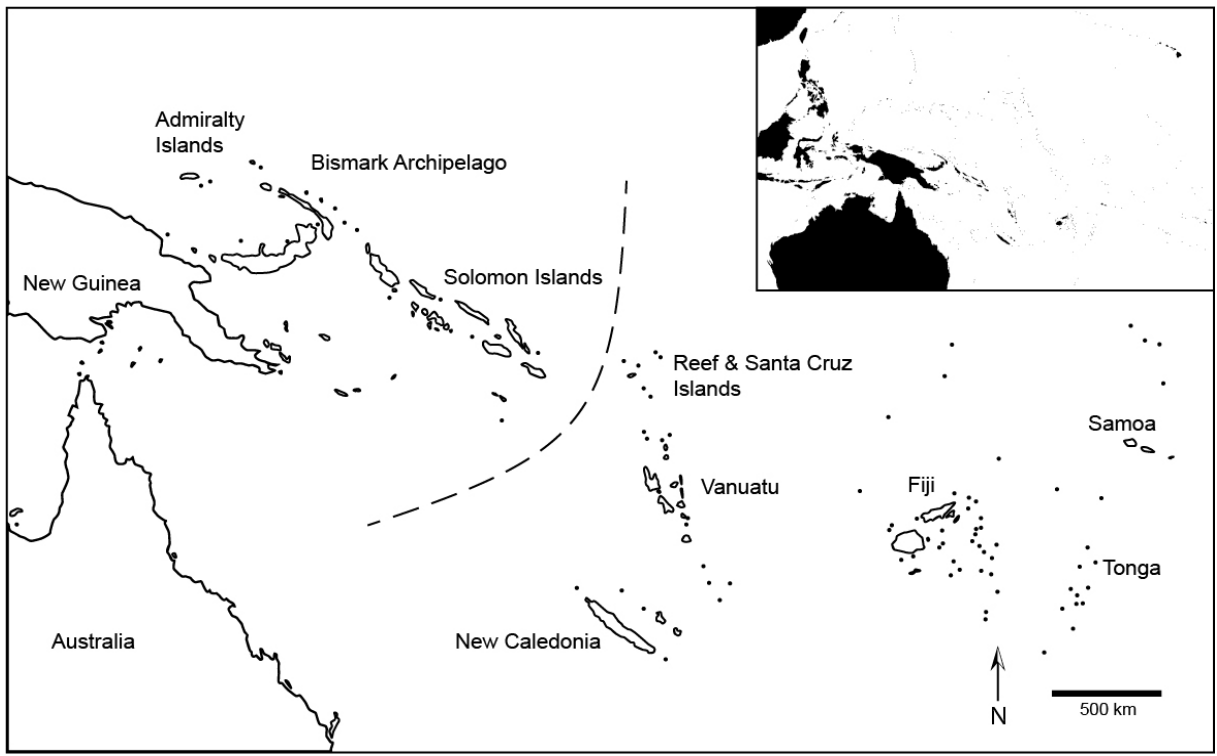

Figure 3: Near and Remote Oceania separated by the dashed line. Australia shown, but not considered Near Oceania.